\documentclass[12pt]{scrartcl}
\usepackage{algorithm2e}
\usepackage{amsmath}
\usepackage{amsfonts}
\usepackage{bm}
\usepackage{bbm}
\usepackage{caption}
\usepackage{graphicx}
\usepackage{natbib}
\usepackage{multirow}
\usepackage{placeins}
\usepackage{setspace}

\newcommand{\R}{\mathbb{R}}
\newcommand{\betaf}{\boldsymbol{\beta}}
\newcommand{\epsf}{\boldsymbol{\epsilon}}
\newcommand{\Af}{\bm A}
\newcommand{\Bf}{\mathbf{B}}
\newcommand{\Df}{\mathbf{D}}

\newcommand{\yf}{\mathbf{y}}
\newcommand{\xf}{\mathbf{x}}
\newcommand{\zf}{\mathbf{z}}
\newcommand{\af}{\bm a}
\newcommand{\Gf}{\bm G}
\newcommand{\gf}{\bm g}

\newcommand{\Omegaf}{\boldsymbol{\Omega}}
\newcommand{\muf}{\boldsymbol{\mu}}
\newcommand{\Sigmaf}{\boldsymbol{\Sigma}}

\begin{document}
\captionsetup{format=hang}

\title{Beyond unimodal regression:\\ modelling multimodality with piecewise unimodal regression or deconvolution models}
\author{Claudia K\"ollmann, Katja Ickstadt and Roland Fried\\$ $ \\Faculty of Statistics, TU Dortmund University\\ 44221 Dortmund, Germany \\koellmann@statistik.tu-dortmund.de}
\maketitle
\onehalfspacing
\section*{Abstract}
Shape constraints enable us to reflect prior knowledge in regression settings. A unimodality constraint, for example, can describe the frequent case of a first increasing and then decreasing intensity. Yet, data shapes often exhibit multiple modes. Therefore, we go beyond unimodal regression and propose modelling multimodality with piecewise unimodal regression or with deconvolution models based on unimodal peak shapes. Usefulness of unimodal regression and its multimodal extensions is demonstrated within three applications areas: marine biology, astroparticle physics and breath gas analysis. Despite this diversity, valuable results are obtained in each application. This encourages the use of these methods in other areas as well.
\newpage
\section{Introduction}\label{sec:intro}
Unimodal regression as a type of nonparametric shape constrained regression is a suitable choice in regression problems when the prior information about the underlying relationship between predictor and response is vague, but when it is (almost) certain that the response first increases and then decreases with higher values of the predictor. The first approach to unimodal regression was a pointwise nonparametric estimation procedure by \citet{Fri1986}. The idea is to estimate two monotonic regressions and find the best location for the transition via the least squares criterion. This has the drawback of possible discontinuity at the mode. The unimodal smoothing approach by \citet{Eilers2005} imposes log-concavity on spline functions using a suitable penalty on the spline coefficients. However, unimodality is not equivalent to log-concavity and caution should be paid if the underlying shape is unimodal, but not log-concave, for example, if the shape is "heavy-tailed".\\
Another semi-parametric spline regression approach to unimodal regression was derived in \citet{kbi2014} and its usefulness in dose-response analysis was demonstrated. Simulations indicated that it is advantageous in comparison to both parametric and nonparametric competitors. The method is based on the fact that, using the B-spline basis, a spline can be restricted to be unimodal by choosing a unimodal sequence of B-spline coefficients with a fixed mode. The use of spline functions guarantees continuity of the fit and smoothness can be achieved by using penalized spline regression as also shown in \citet{kbi2014}. Picking up the idea of \citet{Fri1986}, the mode is selected from all possibilities based on the least squares criterion. This increases computation time, but the possibility to fix the mode can also be advantageous (cf. deconvolution with varying peak shapes in Section~\ref{sec:mult}).\\
However, the prior knowledge in different applications has various degrees of complexity since data shapes may vary from piecewise unimodal relationships to accumulations of identically or even diversely shaped unimodal functions. In this article we argue that unimodal regression is also useful in situations where the relationship between two variables is not unimodal, but multimodal. To be more explicit, by multimodal data we mean observations from a predictor variable $X$ and a response variable $Y$ whose functional relationship has several modes (local maxima). Such multimodality can have different reasons and thus, one can take different approaches when modelling multimodal data. For example, one might observe a series of well separated unimodal responses. For such data piecewise unimodal regression is appropriate. Another reason might be the presence of two or more subpopulations to which the observed entities belong with certain probabilities and where the response variable follows a different unimodal relationship in each subpopulation. A mixture of regression models is able to capture such a structure. Finally, it is also possible that one observes a global response from a series of overlapping and accumulating unimodal processes. In this case, a deconvolution model is needed.\\
This article concentrates on piecewise unimodal regression and deconvolution models, which assume observations from a homogeneous population, and the mixture regression is mentioned just for completeness \citep[for an implementation of the latter, see, e.g., the R package \texttt{flexmix},][]{Gruen2008}. While the application of unimodal regression to several pieces of a data set is quite straightforward and will be briefly described in Section~\ref{sec:piece}, representatives of the class of deconvolution models are very diverse and numerous. This statement still remains valid, if we make the assumption of linear convolution, that is, confining to cases where the observed global response is a linear combination of an unknown number of unobserved input processes (henceforth referred to as "peaks"), i.e., the response vector $\yf$ can be written as $$\yf=\Gf \af + \boldsymbol{\epsilon},$$ where $\af$ are the coefficients of the linear combination, $\boldsymbol{\epsilon}$ the measurement errors and the columns of $\Gf$ describe the peak shapes. The task is then to deconvolve the observed multimodal signal into the single peaks, that is, to estimate both the matrix $\Gf$ and the vector $\af$. This deconvolution problem mathematically belongs to the class of inverse problems and is also known as ``blind source separation'' in signal processing. Since numerous different combinations of $\Gf$ and $\af$ can explain the output equally well, the problem is ill-conditioned and the respective models are underdetermined. To arrive at reasonable solutions nevertheless, various deconvolution algorithms have been proposed which mostly use iterative schemes, such as the Expectation-Maximization (EM) algorithm \citep[see][for an overview]{rooi2011}.\\
In signal processing or chemometrics such approaches are often called blind source separation techniques. A widely used one is MCR, multivariate curve resolution, where ``multivariate'' refers to the fact that usually several contiguous output vectors are obtained and the algorithms are designed for direct application to the composite data matrix with model $\mathbf{Y}=\Gf \Af + \mathbf{E}$. Different MCR methods are, for example, given in \citet{Tauler1995}, \citet{Pomareda2010} and \citet{Oller2015}. A drawback of those approaches is that the number of peaks is not estimated within the model, but by other methods previous to the actual analysis.\\
Bad conditioning and underdetermination can generally be addressed by regularization and the use of constraints. Using positivity constraints in MCR, for example, leads to the so-called non-negative matrix factorization \citep[cf.][]{Pomareda2010}. Other examples of constraints are unimodality or sparsity, the latter one being imposed by regularization in form of a LASSO/$L_1$-penalty \citep[see, for example MCR-LASSO by][]{Pomareda2010} or by an $L_0$-penalty \citep{rooi2011}. We follow the latter authors and use the $L_0$-penalty, which allows for estimation of the number of peaks simultaneous to the other model parameters. Since their original model is restricted to pointwise estimation of identically shaped peaks, we make some enhancements, for example, enabling the estimation of a functional description of identically shaped peaks. The most interesting development, however, is our combination of the deconvolution model using $L_0$-penalty with the "additive unimodal regression model" (see Equation~(\ref{eq:add})). The combined approach enables sparse deconvolution of multimodal data using differently shaped unimodal spline components.\\
Apart from using penalties for regularization, the unimodality constraint is of great importance here. We will impose unimodality on spline functions throughout this article by using the penalized unimodal spline regression by \citet{kbi2014}. In many situations log-concave spline smoothing suggested by \citet{Eilers2005} can be used as well, but regarding the deconvolution model with varying peak shapes in Section~\ref{sec:mult} the approach by \citet{kbi2014} is advantageous.\\
In the following, we will demonstrate with three real data examples from different application areas that unimodal regression in general is useful for modelling multimodality. The applications are increasing in complexity as they vary from piecewise unimodal to convolutions of identically or even diversely shaped unimodal functions.
\subsection{Analysis of dive phases of marine animals}\label{sec:diveintro}
The first field of application is the analysis of diving behaviour of marine animals. Time-depth-recorders (TDRs) measure the diving depth of marine animals such as seals or whales. The resulting data sets may contain measurements over several days at regular sampling frequencies. In the case of marine mammals, the animals repeatedly perform dives from the water surface down to various depths to find food and for other activities. An excerpt from such a TDR data set, taken from the R package \texttt{diveMove} \citep[cf.][]{diveMove2007,Luq2011}, is shown in Figure \ref{fig:data}A. Marine biologists are interested, among other things, in the detection of phases within a dive that correspond to different behaviours \citep[see e.g.][]{Hals2007}. The last-mentioned authors claim that there is need for objective and automated categorisation of the diving behaviour and develop a Matlab program classifying diving depth data with a set of prespecified criteria. This approach does not consider measurement error, which can be accounted for by modelling the dives statistically. This is, for example, realized in version 1.4.1 of the R package \texttt{diveMove} by fitting multiple smoothing splines. Afterwards, the derivative of the fitted splines is used to divide the dives into different phases like descent and ascent. In this article we show that using piecewise unimodal regression splines is advantageous for this purpose. Since the animal definitely has to come back to the surface to draw breath, in explicit, since the dives do not overlap, the time series can be modelled by \emph{piecewise unimodality}.
\subsection{Astroparticle physics data analysis}\label{sec:factintro}
\begin{figure}
\centering
		\includegraphics[width=0.95\textwidth]{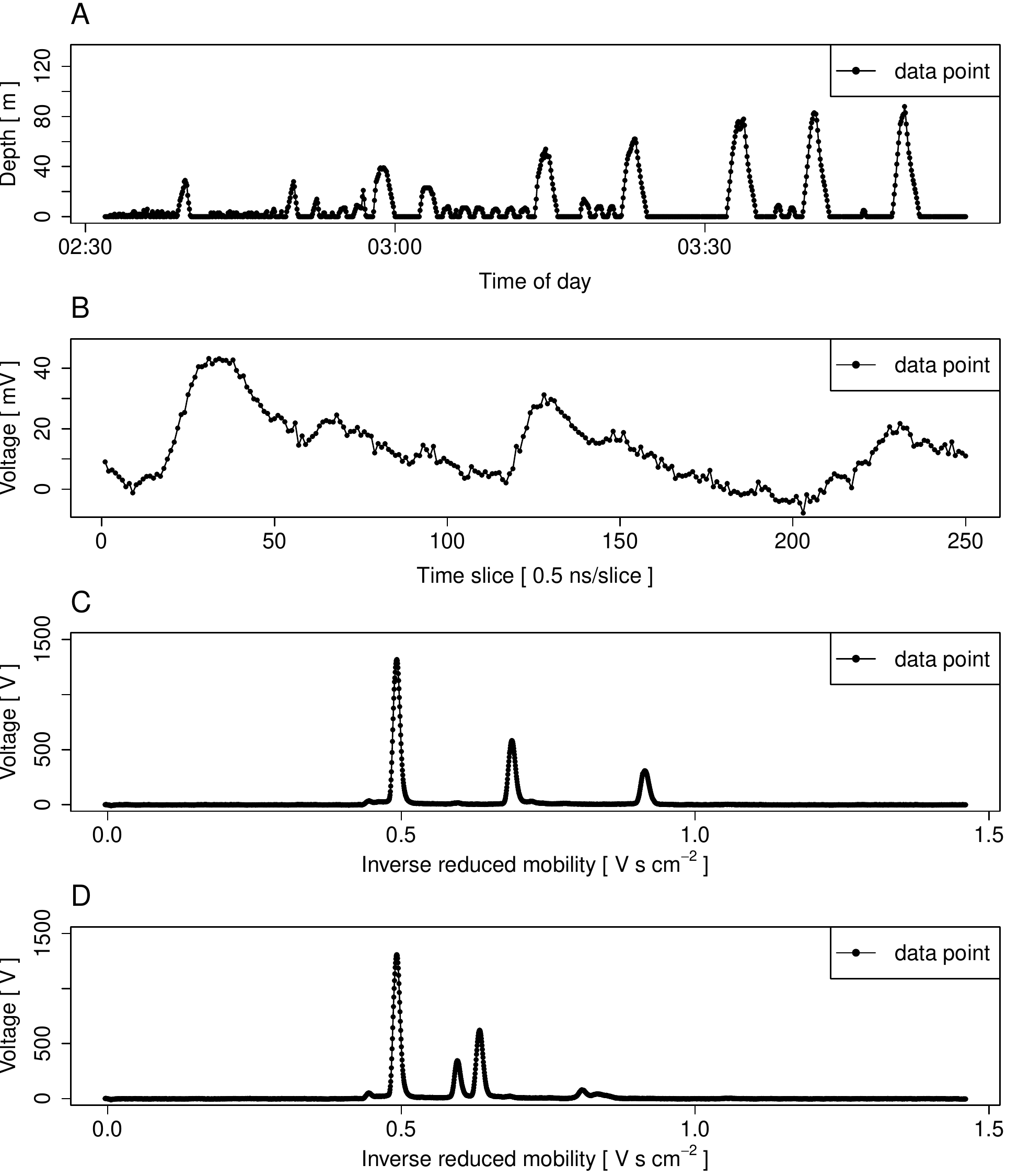}
	  \caption{\textbf{Example data sets.} (A) shows an excerpt from data set \emph{divesTDR} \citep[R package \emph{diveMove},][]{diveMove2007}. It displays the diving depth [in m] of a marine animal, which was recorded every five seconds between 02:31:55 a.m.~and 03:55:15 a.m.~on January 6th 2002. (B) is an example of a FACT time series. The x-axis is the number of the time slice or sample, where the slices are about 0.5 ns wide. The y-axis gives the measured voltages in millivolt [mV]. (C) and (D) show spectrum A and B of the IMS example data set. The x-axis is inverse reduced mobility [V s cm$^{-2}$], a transformation of drift time. The y-axis gives the voltage measured in volt [V].}\label{fig:data}
\end{figure}

The second application comes from astroparticle physics. The First G-APD Cherenkov Telescope \citep[FACT; see][]{fact2013,Bil2012} is used by astroparticle physicists to detect cosmic rays. These cosmic rays induce light flashes in the earth's atmosphere, which can be used to calculate the primary particle's properties. The camera of the telescope has several pixels and each pixel collects a signal, that is, a time series of measured voltages during one event. See Figure \ref{fig:data}B for an example. Each photon hitting a camera pixel causes a change in the signal, which can be described by a unimodal loading curve with an amplitude of approximately 10 mV \citep[][]{fact2013}. The aim is to detect the arrival times and numbers of photons to draw conclusions about the type of the triggering particle (gamma or hadron). A good overall fit is of interest, too, since the integral over the signal is used in subsequent analyses. The shape of the signal is similar to that of a loading and unloading capacitor and thus, physicists have suggested a parametric wave form for the change in the voltage due to the arrival of one or more photons. When $n_p$ photons arrive at time $t_0$ this wave form is given by
\begin{eqnarray}
U(t)=\gamma + n_p\cdot U_0\cdot \mathcal{I}_{[t_0,\infty)}(t) \cdot \left(1-e^{-\frac{t-t_0}{\xi_1}}\right)e^{-\frac{t-t_0}{\xi_2}}, \label{eq:wave}
\end{eqnarray}
where $\mathcal{I}_{\mathcal{A}}(.)$ denotes the indicator function over the set $\mathcal{A}$ and $\gamma$ is the baseline voltage shortly before the photons' arrival \citep[cf.][formula 6.11]{Buss2013}.
Since the telescope has been constructed quite recently, standard methods for the evaluation of measured signals are mostly heuristic and only applied on segments of a signal. Parameter estimates for waves of the form (\ref{eq:wave}) from well-distinguished signals of \emph{single} photons ($n_p=1$) were derived in \citet{Buss2013}. As photons can arrive any time, the measured voltage is an accumulation of several loading curves (each corresponding to one or more photons). This suggests using a \emph{deconvolution model} with accumulated parametric waves for the analysis of a whole time series of one pixel. For the deconvolution model that will be used in Section \ref{sec:fact} the parameters $\gamma$, $n_p$ and $t_0$ can be fixed at certain values. Explicitly, the baseline $\gamma$ can be set to 0 since the deconvolution model already takes care of the fact that the individual waves start at a higher baseline voltage due to the convolution. In addition, the deconvolution model estimates implicitly the arrival time and the number of photons, so that $t_0$ and $n_p$ can be set to 0 and 1, respectively, and each single wave is described by
\begin{eqnarray}
U_{\mbox{single}}(t)= U_0\cdot \mathcal{I}_{[0,\infty)}(t) \cdot \left(1-e^{-\frac{t}{\xi_1}}\right)e^{-\frac{t}{\xi_2}}. \label{eq:wave2}
\end{eqnarray}
The remaining parameters can be estimated, for example, using the least squares criterion, or fixed at the following values derived in \cite{Buss2013} for the data at hand: $\hat{U_0}=17.41$, $\hat{\xi_1}=4.745$, $\hat{\xi_2}=31.81$.
\subsection{Breath gas analysis with ion mobility spectrometry}\label{sec:imsintro}
The third application area is breath gas analysis where ion mobility spectrometry (IMS) coupled with multicapillary columns (MCCs) is used to measure the amount of certain molecules in the air or in exhaled breath. Knowledge about the presence of such molecules and their concentrations can be used for medical purposes, for example, to diagnose lung cancer \citep[cf.][]{West2009}. An IMS-MCC data set is a matrix of measured intensities. Looking only at one row or one column at a time, the intensities are time series along drift time (rows) or along retention time (columns). In this article we focus only on the observed intensities along the drift time, which are called spectrum. Typically, the intensities in a spectrum fluctuate around zero and exhibit few peaks, see also Figures \ref{fig:data}C and \ref{fig:data}D. At least one peak is always present at about 0.5 and does not carry information about the analyte: the so-called reaction ion peak. The other peak locations and their amplitudes provide information about the presence of different molecule types. Since the IMS technology is getting more and more miniaturized and feasible for mobile use, there is also need for suitable analysis methods, which a) work automatically, in explicit, without an expert, and b) process the data online (e.g.~analysing one spectrum during measurement of the next).\\
One of the main steps in IMS analysis is peak extraction, that is, data reduction in a way that every peak is described by a number of parameters, which reproduce the characteristics of a spectrum as closely as possible, for example, with location and amplitude. Currently, this is done manually by experts using interactive visualization software. Subsequent analyses use the parameters of all spectra to identify different molecules in the analyte or to classify samples into subgroups like healthy or not \citep[see also][]{Hausch2013}. In this article we only focus on the peak extraction step and aim at modelling a single spectrum at a fixed retention time statistically.\\
For whole IMS-MCC data matrices \citet{kop2012} used a 2-dimensional mixture model with a background component and several peak components, where each of the latter ones is based on the product of two shifted inverse Gaussian distributions, one for drift time and one for retention time. In a more recent approach \citet{kop2015} again use shifted inverse Gaussian distributions to model each peak in one spectrum (row) of IMS-MCC data sets, which are combined to 2-dimensional models in a subsequent step. This means that a peak is modelled with only a small number of parameters in both cases. Other approaches for modelling IMS or similar data \citep[see, e.g.][]{Vogt2009,Boed2009,Ross2006} use different distributions, also with very few parameters. Although one of the aims is data reduction and the distributions allow, for example, for skewness, this representation of the data might be too restrictive regarding the shape of a peak. Additionally, those approaches have the disadvantage that a distribution family has to be specified prior to the analysis. Since it is known that each peak is a unimodal function of the drift time, we propose instead to describe a spectrum with multiple unimodal spline functions. Thus, we will consider \emph{piecewise} and \emph{deconvolution} models in this application.\\
The remainder of the article is organized as follows: Section \ref{sec:meth} gives an overview of the statistical models and methods. We first describe the unimodal spline regression approach by \citet{kbi2014} before we extend it to models for multimodal data: piecewise unimodal regression and deconvolution models based on unimodal peak shapes. The three subsections of Section \ref{sec:res} describe how the data sets from the different application areas were analysed and which results were gained. A summary and discussion is included in Section \ref{sec:disc}.
\section{Methods}\label{sec:meth}
This section aims at presenting methodology which enables the handling of a broad spectrum of applications with multimodal data. Thus, we provide an overview of several approaches and give recommendations on the method of choice in different situations. A common feature of the methods is that the response variable is modelled using several unimodal functions, which will also be referred to as "peaks".\\
There are applications where knowledge about the shape of the individual peaks exists. The shape might either be known exactly or can be described by a parametric function. If this is not the case, more sophisticated estimation methods for the peak shape are required. In this article we employ the unimodal regression approach proposed in \citet{kbi2014}. This semi-parametric spline regression approach, which enables the fitting of very diverse peak shapes, is described in Section~\ref{sec:uni}. The multimodal regression approaches generalizing it are presented in Section~\ref{sec:mult} and recommendations on the model choice are given in Section~\ref{sec:recomm}.\\
We first introduce some notational aspects. Let $(x_i,y_i) \in [a,b] \times \R$ be pairs of observations that may be modelled by \begin{eqnarray} Y_i=f(x_i) + \epsilon_i,\ i=1,\hdots,n, \label{eq:model} \end{eqnarray}
where we assume the measurement errors $\epsilon_i \underset{i.i.d.}{\sim} \mathcal{N}(0,\sigma^2)$ throughout the article, while the form of the function $f$ will be determined by the different methods.\\
Let $\mathcal{T}=(\tau_j)_{-k}^{q+k+1}$ be the sequence of knots of a B-spline basis for the interval $[a,b]$. That is, the basis consists of $d=q+k+1$ (normalized) B-spline basis functions, $N_{j,k+1}(x)$, of degree $k \ge 1$ with knots $\tau_j,\hdots,\tau_{j+k+1}$. For each $x \in [a,b]$ the function values are given by the recursion formulae $N_{j,1}(x)=\mathcal{I}_{[\tau_j,\tau_{j+1})}(x),$ $N_{j,k+1}(x)=\frac{x-\tau_j}{\tau_{j+k}-\tau_j}N_{j,k}(x) + \frac{\tau_{j+k+1} - x}{\tau_{j+k+1}-\tau_{j+1}} N_{j+1,k}(x)$ for $j=-k,\hdots,q$. Since those functions form a basis of the space of spline functions on $[a,b]$ of degree $k$ with knots $\mathcal{T}$, every spline function $s$ in this space can be written uniquely as $s(x)=\sum_{j=-k}^q \beta_j N_{j,k+1}(x)$ with so-called B-spline coefficients $\beta_{-k},\hdots,\beta_q$. For details on spline functions and spline regression see, for example, \citet{Die1993}.\\ 
We denote with $\displaystyle \mathcal{N}_M(\muf,\Sigmaf)$ a multivariate normal distribution with mean $\muf \in \R^d$ and covariance matrix $\Sigmaf \in \R^{d \times d}$ truncated to the set $M \subset \R^d$. Furthermore, we define $\yf:=(y_1,\hdots,y_n)' \in \R^n$, $\epsf:=(\epsilon_{1},\hdots,\epsilon_n)' \in \R^n$, $\betaf:=(\beta_{-k},\hdots,\beta_q)' \in \R^d$ and 
$$\Bf:=\begin{pmatrix} N_{-k,k+1}(x_1)& \cdots & N_{q,k+1}(x_1)\\ \vdots & & \vdots\\ N_{-k,k+1}(x_n)& \cdots & N_{q,k+1}(x_n) \end{pmatrix} \in \R^{n \times d},$$
is the matrix of B-spline basis functions evaluated at the observations $x_1,\hdots,x_n$.\\
In this article we use cubic (degree $k=3$) splines for all analyses and we suppose that an adequate estimate of the error variance $\sigma^2 \ge 0$ is available prior to model fitting.
\subsection{Unimodal regression}\label{sec:uni}
The building block of each model used for analysing the introduced data types is the penalized unimodal spline regression proposed in \citet{kbi2014}. Here, the function $f$ from model (\ref{eq:model}) is given by a unimodal spline function $s(x)$.\\
The shape constraint is imposed by finding the spline $s(x) = \sum_{j=-k}^q \beta_j N_{j,k+1}(x)$ or respectively the corresponding coefficient vector $\betaf$, that minimizes a penalized residual sum of squares criterion subject to the condition 
$\betaf \in \mathcal{S}_m=\{\betaf \in \R^d\ |\ \beta_i \ge \beta_{i-1} \ \forall \ i \le m \text{ and } \beta_i \le \beta_{i-1} \ \forall \ i > m \}$.\\
The penalized residual sum of squares criterion is given by
\begin{eqnarray}
\mathit{pRSS} & = & \frac{1}{\sigma^2}\sum_{i=1}^n \left\{y_i - \sum_{j=-k}^q \beta_j N_{j,k+1}(x_i) \right\}^2 + \lambda \left\|\Omegaf^{\frac{1}{2}}(\betaf - \betaf_0) \right\|^2_2\notag \\ 
& = & \frac{1}{\sigma^2}\left\|(\yf - \Bf\betaf) \right\|^2_2 + \lambda \left\|\Omegaf^{\frac{1}{2}}(\betaf - \betaf_0) \right\|^2_2, \label{eq:prss}
\end{eqnarray}
where $\Omegaf \in \R^{d \times d}$ is a matrix of penalty coefficients and $\betaf_0 \in \R^d$ is a vector of constants. Choosing a large number of knot positions and placing such a penalty on the B-spline coefficients leads us to a compromise between overfitting and underfitting. The choice of $\Omegaf$ and $\betaf_0$ determines the form of the penalty, for example, one can use differences of the B-spline coefficients as described in \citet{EilMar1996} or penalization against parametric functions as proposed in \citet{kbi2014}. In case of the IMS data, we use a zero order difference penalty, more typically called "ridge penalty", that is, we penalize against a constant zero function since the intensities are very small over most of the x-axis. This penalty is obtained when $\Omegaf$ is the $d \times d$ identity matrix and $\betaf_0 = \mathbf{0}$. In the diving depth example, each unimodal piece is fitted with a second order difference penalty to introduce smoothness, that is, $\Omegaf=\Df_2'\Df_2$ ($\Df_2 \in \R^{(d-2) \times d}$ being the matrix of second order differences) and $\betaf_0 = \mathbf{0}$.\\
The tuning parameter $\lambda$ in (\ref{eq:prss}) can be chosen via restricted maximum likelihood estimation (REML) or approximate REML, respectively, as described in \citet{kbi2014}. This means that the coefficients are viewed as random with prior $\betaf | \lambda   \sim  \mathcal{N}_{\mathcal{S}_m}(\betaf_0,\frac{1}{\lambda} \Omegaf^{-1})$ or $\betaf | \lambda   \sim  \mathcal{N}_{\R^d}(\betaf_0,\frac{1}{\lambda} \Omegaf^{-1})$, respectively, and integrated out of the joint likelihood to obtain the restricted likelihood of $\lambda$. The tuning parameter that maximizes the restricted likelihood is selected.\\
Since the mode $m$ of the coefficients is unknown when fitting a unimodal spline to data, the minimization is performed for each possible choice of $m$ and a decision is made subject to the residual sum of squares criterion.
\subsection{Multimodal regression approaches}\label{sec:mult}
As already mentioned in the introduction, multimodality can have different sources and thus there is need for different modelling approaches. In the following, the piecewise unimodal regression and deconvolution models based on identical and diverse unimodal peak shapes are presented.
\subsubsection*{Piecewise unimodal regression}\label{sec:piece}
A simple approach for modelling multimodal data is piecewise unimodal regression, that is, dividing the x-axis (heuristically) between each pair of modes and fitting separate unimodal splines. The function $f$ from model (\ref{eq:model}) can be written as $$f(x)= \sum_{k=1}^K s_k(x)\mathcal{I}_{I_k}(x),$$
where $I_1,\hdots,I_K \subset [a,b]$ are $K$ intervals corresponding to the x-axis' pieces and $s_k$ are unimodal spline functions on the respective intervals. Depending on the application, there are different ways to determine these intervals, for example, using a threshold.\\
This model implies the assumption that the underlying process, which generates the observations, is also divisible in some respects. We will refer to this modelling approach as pUniReg.
\subsubsection*{Deconvolution with identical peak shapes}\label{sec:ident}
Other multimodal regression approaches, so-called deconvolution models, describe the observations as a convolution of peaks. As opposed to the former approach, they allow modelling of overlapping peaks which accumulate at the overlap to the observed values, but can also be applied, when no overlap is present. In this paper, we examine only linear convolution.\\
Let us first assume that all peaks have the same basic shape and each peak is a scaled version of it. This can be expressed by writing the function $f$ from model (\ref{eq:model}) as
\begin{eqnarray*}
f(x_i)= \sum_{j=1}^{n_g} g_j a_{i-j},\ i = 1,\hdots, n,
\end{eqnarray*}
where $\gf=(g_1,\hdots,g_{n_g})'$ is the known (pointwise) peak shape and the vector $\af=(a_{-n_g+1},\hdots,a_{n-1})'\in \R^{n+n_g-1}$ holds the so-called input pulses, which describe the number of peaks (given by the number of values $a_j \neq 0$), their locations (given by index $j$) and heights (given by the actual value of $a_j$). The number $n_g$ of $x$-points, for which the peak shape is given, is usually smaller than $n$ and thus, the individual peaks do not span over the whole range of $x$-values. This model was introduced by \citet{rooi2011} and it was shown that it can also be reformulated as a typical linear regression model:
\begin{eqnarray*}
\yf=\Gf \af + \bm{\epsilon},
\end{eqnarray*}
where the convolution matrix $\Gf \in \R^{n \times (n+n_g-1)}$ holds shifted copies of the peak shape $\gf$ in its columns (compare also to the general linear deconvolution model in Section~\ref{sec:intro}).\\
If the peak shape $\gf$ is known, the least squares estimate of $\af$ is given by $\hat{\af}=(\Gf'\Gf)^{-1}\Gf\yf$, but the columns of $\Gf$ are highly correlated, which leads to illposedness. This problem was already described by \citet{rooi2011} and the authors propose to use regularization with an $L_0$-penalty on $\af$, that is, using the objective function $\left\|\yf - \Gf\af \right\|^2_2 + \kappa \sum_{j} |a_j|^0$. The regularized estimate for $\af$ is found by minimizing the objective function using an iterative procedure. Since the penalty factor $\sum_{j} |a_j|^0$ is essentially the number of peaks, the regularized estimation favours sparse models with only few peaks. The higher the tuning parameter $\kappa$, the fewer the peaks. \citet{rooi2011} propose to choose the tuning parameter $\kappa$ by visual inspection. They also describe an additional, asymmetric penalty on $\af$ that renders the estimated input pulses positive. Altogether, the procedure is able to estimate the number of peaks, their locations and heights simultaneously. We will refer to this modelling approach as $L_0$-deco.\\
Additionally, \citet{rooi2011} present an approach for cases where the peak shape $\gf$ is unknown, which is called "blind deconvolution". The idea is, starting with an initial pointwise peak shape $\gf^{(0)}$, to iterate between estimation of $\af$ and $\gf$. Given an (interim) estimate of $\af$ a new estimate of $\gf$ can be found in three different ways. The first one is described in \citet{rooi2011} and produces a pointwise estimate of $\gf$. It can be found using the reformulated model $\yf=\Af\gf + \bm{\epsilon}$. Here, $\Af$ holds shifted copies of $\hat{\af}$ in its columns and the pointwise least-squares estimate is given by $\hat{\gf}=(\Af'\Af)^{-1}\Af\yf$. It is also possible to use a smoothness penalty (differences penalty or unimodal smoother) on the entries of $\gf$ \citep[cf.][]{rooi2014}. We will refer to this approach as pointwise $L_0$-deco.\\
In this article we want to estimate not only a smooth pointwise description of each peak, but also a continuous functional peak shape. \citet{rooi2011} already mentioned the possibility to use spline functions for this purpose. Here, we give details for a slightly more general approach, where the peak shape is given by a function $g(x|\betaf)$ parameterized by vector $\betaf$. Either the peak shape is known to (approximately) follow a unimodal parametric function, where the parameters in $\betaf$ are usually few and nicely interpretable. Or the function $g$ can be a semi-parametric unimodal spline function, if there is no prior information about the peak shape. In both cases, function $f$ from model (\ref{eq:model}) can be written as $f(x_i)=\sum_{j=1}^{n_g} g(x_j|\betaf)a_{i-j},\ i=1,\hdots,n$ and the following estimation procedure can be applied:\\
With initial parameter vector $\betaf^{(0)}$ and peak shape $\gf^{(0)}=(g(x_1|\betaf^{(0)}),\hdots,g(x_{n_g}|\betaf^{(0)}))'$ respectively, it is again possible to iterate between estimation of $\af$ and $\betaf$. In the $k$-th iteration, estimation of $\af$ is done as described above using the peak shape $\hat{\gf}^{(k-1)}=(g(x_1|\hat{\betaf}^{(k-1)}),\hdots,g(x_{n_g}|\hat{\betaf}^{(k-1)}))'$, where $\hat{\betaf}^{(k-1)}$ is the current estimate of $\betaf$. The parameter vector $\betaf$ can be estimated using the least squares method, that is, minimizing $||\yf - \Af\gf||^2 = ||\yf - \Af (g(x_1|\betaf),\hdots,g(x_{n_g}|\betaf))'||^2$ with respect to $\betaf$ or with unimodal regression. These two blind deconvolution approaches will be referred to as parametric $L_0$-deco and unimodal $L_0$-deco. Except for the initial guess, no information about the basic peak shape is needed for the unimodal $L_0$-deco approach. All blind deconvolution approaches simultaneously obtain estimates of a basic peak shape, the number, locations and heights of the peaks. The parametric and unimodal $L_0$-deco have the advantage of a functional shape description.
\subsubsection*{Deconvolution with diverse peak shapes}
While there are several applications in which the assumption of the same shape for all peaks is very plausible (for example, the FACT data), there are also situations where the observed signal is a convolution of peaks with different shapes (for example, in IMS data). \citet{Eilers2005} used sums of log-concave smoothing splines for such a deconvolution task. More generally speaking, an appropriate representative of deconvolution models in this situation is an additive model, that describes the observations as convolution of $L$ different peak shapes. In this model the function $f$ from Equation (\ref{eq:model}) is given by
\begin{eqnarray} f(x_i) = \alpha + \sum_{\ell=1}^L g_\ell(x_i), \label{eq:add}
\end{eqnarray}
where $\alpha$ is an intercept and each $g_\ell(x)$ is a unimodal function describing one of the peaks and can be evaluated over the whole range of the $x$-observations. \citet{Eilers2005} used log-concave smoothing splines for each $g_\ell$, but of course each of these functions can be described by a parametric model (with different parameter values) or by unimodal spline regression (with different parameters and modes) as well. For the latter choice we have $g_\ell(x)=\sum_{j=-k}^q \beta_{\ell,j} N_{j,k+1}(x)$ with coefficients $\beta_{\ell,-k} \le \hdots \le \beta_{\ell,m_\ell -1}\le \beta_{\ell,m_\ell} \ge \beta_{\ell,m_\ell+1} \ge \hdots \ge \beta_{\ell,q}$.\\
Additive models can be fitted using the so-called backfitting algorithm \citep[cf.][]{Hast2009}, which is given by
\begin{enumerate}
\item Initialize $\hat{\alpha} = \frac{1}{n}\sum_{i=1}^n y_i$ and $\hat{g}_\ell(x) \equiv 0 \ \forall \ell$.
\item For $\ell=1,\hdots,L$: calculate $\hat{g}_\ell$ from data $(x_i,\tilde{y}_i)$ with $\tilde{y}_i=y_i-\hat{\alpha} - \sum_{k \neq \ell}\hat{g}_k(x_i),$ $i=1,\hdots,n.$
\item Center the function estimates around zero: $\hat{g}_\ell=\hat{g}_\ell - \frac{1}{n}\sum_{i=1}^n\hat{g}_\ell(x_i)$.
\item Repeat steps 2 and 3 until convergence.
\end{enumerate}
In contrast to commonly applied additive models \citep[see e.g.][]{Hast2009} we have only one regressor that is used in all components. Thus, the number $L$ of components is not simply the number of regressors. Sometimes the specific application might enforce a fixed number of components or it can be determined with the help of a model selection criterion, for example Akaike's information criterion (AIC). 
We will refer to this modelling approach as addUniReg.\\
In principle, the addUniReg model is applicable in all mentioned situations, since it is the most general model. This flexibility comes, for example, at the cost of higher computation times, because each component is estimated with unimodal regression (involving determination of the mode by trying all possibilities) and the number of components can only be determined by fitting several models and making a choice based on AIC. Thus, the $L_0$-deconvolution model is preferable, since it simultaneously estimates the number, the locations and heights of peaks. Yet, it cannot cope with different peak shapes.\\
In \citet{rooi2014} the authors already discussed the fact that the peak shapes might vary over time, or more generally speaking, that the peaks might have different shapes. Two concepts were proposed to tackle such problems with $L_0$-deconvolution: finding a transformation of the x-axis such that the peak shape is constant or estimating the matrix $\Gf$ (holding the varying peak shapes in its columns) as a smooth two-dimensional surface. For the first proposal, knowledge about the way in which the peak shape changes over time would be required and this is not the case for the applications we have in mind (for example, IMS data). The here presented approach is in line with the second suggestion, although the surface will only be smooth in one of its directions (each column holds a smooth peak shape). It can be seen as a combination of the $L_0$-deconvolution model and the additive unimodal regression.\\
Suppose we have response values $\yf$ with a baseline fluctuating around zero and several peaks with maximum peak height equal to one (data can easily be transformed to fulfil these criteria). For this model the function $f$ from Equation (\ref{eq:model}) can be written as follows:
\begin{eqnarray*}
f(x_i)=\sum_{j=1}^d g_{j}(x_i) a_j,
\end{eqnarray*}
where $g_{j}(.)$ is a spline function with $d$ B-spline coefficients, which have a fixed mode at $j \in\{1,\hdots,d\}$, and $a_j$ is the input pulse corresponding to the $j-th$ peak. In explicit, this is an additive model of $d$ spline components with different mode locations, scaled by the input pulses. The matrix of varying peak shapes $\Gf$ then consists of the values $g_{ij}=g_{j}(x_i), i=1\hdots,n, j=1,\hdots,d$. The input pulses $\af$ have a slightly different role in this model compared to the original deconvolution model. Each input pulse corresponds to one of at most $d$ peaks, where $d$ is the number of B-spline coefficients which is usually much smaller than $n$, but also larger than the number of existent peaks. Thereby the index $j$ does not correspond to observation $x_j$ and the peak locations cannot be derived directly, but the model still provides simultaneous estimation of number and heights of the peaks and additionally estimates the different peak shapes. The ability to estimate spline functions with a fixed mode is very essential here. This could not be achieved as easily with other unimodal regression approaches like, for example, log-concave smoothing by \citet{Eilers2005}.\\
The model can be fitted by iteratively estimating the $g_j$ with steps similar to the backfitting algorithm for the additive model and estimating the input pulses $\af$ using the $L_0$-penalty. After each spline estimation the fitted values are transformed to the interval [0,1] to ensure interpretability (the input pulses describe the heights of the peaks) and identifiability. The estimation procedure is illustrated in Algorithm ~\ref{alg:deco} in the Appendix. We will refer to this model as varying $L_0$-deco.
\subsection{Applicability of the model types}\label{sec:recomm}
Table \ref{tab:choice} gives an overview of the different data situations where the proposed approaches to multimodal regression are applicable. Depending on the shape of the peaks and their overlap the table states the applicable models and our recommendation (in bold).\\ 
In principle, the addUniReg and varying $L_0$-deco models are applicable in all mentioned situations since they cope with the most general situation of overlapping peaks with diverse peak shapes. For this flexibility one has to pay the prize of higher computation times (each peak is estimated over the whole range of the $x$-values), especially for the addUniReg model, because the number of components can only be determined by fitting several models and making a choice based on AIC. Thus, both approaches should only be considered, when there are overlapping peaks with diverse shapes. The varying $L_0$-deco model is to be preferred, if the number of peaks is unknown. If there are differently shaped, but non-overlapping peaks, the simpler pUniReg approach is sufficient, which is in principle also "downwards compatible" to situations with identical peak shapes. Nevertheless, the method of choice when all peaks have the same shape and regardless if there is overlap or not, is the $L_0$-deconvolution model, since it simultaneously estimates the number, the locations and heights of peaks. When the peak shape is unknown, blind deconvolution can be used and we propose to apply the advanced versions, parametric and unimodal $L_0$-deco, instead of pointwise $L_0$-deco to obtain smooth function estimates.\\
As the derivatives of the fitted peaks are also of interest in some applications (see, for example, the diving depth data analysis in Section \ref{sec:dive}), it is important to note that derivatives are easily obtained with all approaches that use (unimodal) splines.
\begin{minipage}{\linewidth}
\centering
		\captionsetup{type=table}
	\caption{\textbf{Overview of the proposed multimodal regression approaches and in which situation to use them.} The recommended model is marked by boldface. The abbreviations are as follows: $L_0$-deco: $L_0$-deconvolution model with fixed peak shape, pointwise $L_0$-deco: blind $L_0$-deconvolution model with pointwise peak shape, parametric $L_0$-deco: blind $L_0$-deconvolution model with parametric peak shape, unimodal $L_0$-deco: blind $L_0$-deconvolution model with unimodal peak shape, pUniReg: piecewise unimodal regression, addUniReg: additive unimodal regression, varying $L_0$-deco: blind $L_0$-deconvolution model with diverse unimodal peak shapes.}\label{tab:choice}
	\begin{tabular}{llp{3.8cm}p{3.8cm}}
		& & no overlap & overlap\\
		\cline{2-4}
		\multirow{6}{1.1cm}{peak\newline shapes} & \multicolumn{1}{|l|}{identical, known}  & $\boldsymbol{L_0}$\textbf{-deco}  & \multicolumn{1}{l|}{$\boldsymbol{L_0}$\textbf{-deco}}\\[0.4cm]
& \multicolumn{1}{|l|}{identical, unknown} & \parbox[t]{3.8cm}{pointwise $L_0$-deco\\ parametric $L_0$-deco\\ \textbf{unimodal} $\boldsymbol{L_0}$\textbf{-deco}} & \multicolumn{1}{c|}{\parbox[t]{3.8cm}{pointwise $L_0$-deco\\ parametric $L_0$-deco\\ \textbf{unimodal} $\boldsymbol{L_0}$\textbf{-deco}\\$ $}}\\
																						& \multicolumn{1}{|l|}{diverse, unknown} & \textbf{pUniReg} & \multicolumn{1}{c|}{\parbox[t]{3.8cm}{addUniReg\\ \textbf{varying} $\boldsymbol{L_0}$\textbf{-deco}}}\\
\cline{2-4}
	\end{tabular}
\end{minipage}
\section{Results}\label{sec:res}
In the following the methods presented in Section \ref{sec:meth} are applied to data sets from three application areas: marine biology, astroparticle physics and breath gas analysis. All analyses were performed using R \citep[version 3.3.0;][]{R2016}. The unimodal penalized spline regressions are fitted using function \texttt{unireg} in R package \texttt{uniReg} \citep{unireg2016}. We use approximate REML to reduce the computational burden that arises from (repeatedly) estimating several unimodal regression functions.
\subsection{Analysis of dive phases of marine animals}\label{sec:dive}
An approach to determine dive phases (descent and ascent) in TDR data is implemented in the R package \texttt{diveMove} \citep[cf.][version 1.4.1]{diveMove2007, Luq2011}. The procedure starts with heuristically splitting the diving depth time series into dives using a depth threshold of three meters and fitting a smoothing spline to each dive (cf. Figure \ref{fig:dive2}A). Afterwards the derivative of the smoothing spline is used to identify the descent and ascent phase of each dive. This determination can be problematic since the uniqueness of the turning point depends on the choice of the smoothing parameter. This can be seen in Figure \ref{fig:dive2}B; for a smoothing parameter chosen via data-driven cross-validation the derivative of the smoothing spline is quite wiggly and crosses the interesting region around the zero derivative several times. For a manually chosen (larger) smoothing parameter the derivative gets smoother and the zero line is only crossed once. However, such a manual choice  is subjective and might be a difficult task for users.\\
\begin{minipage}{\linewidth}
\centering
		\captionsetup{type=figure}
  \includegraphics[width=0.8\textwidth]{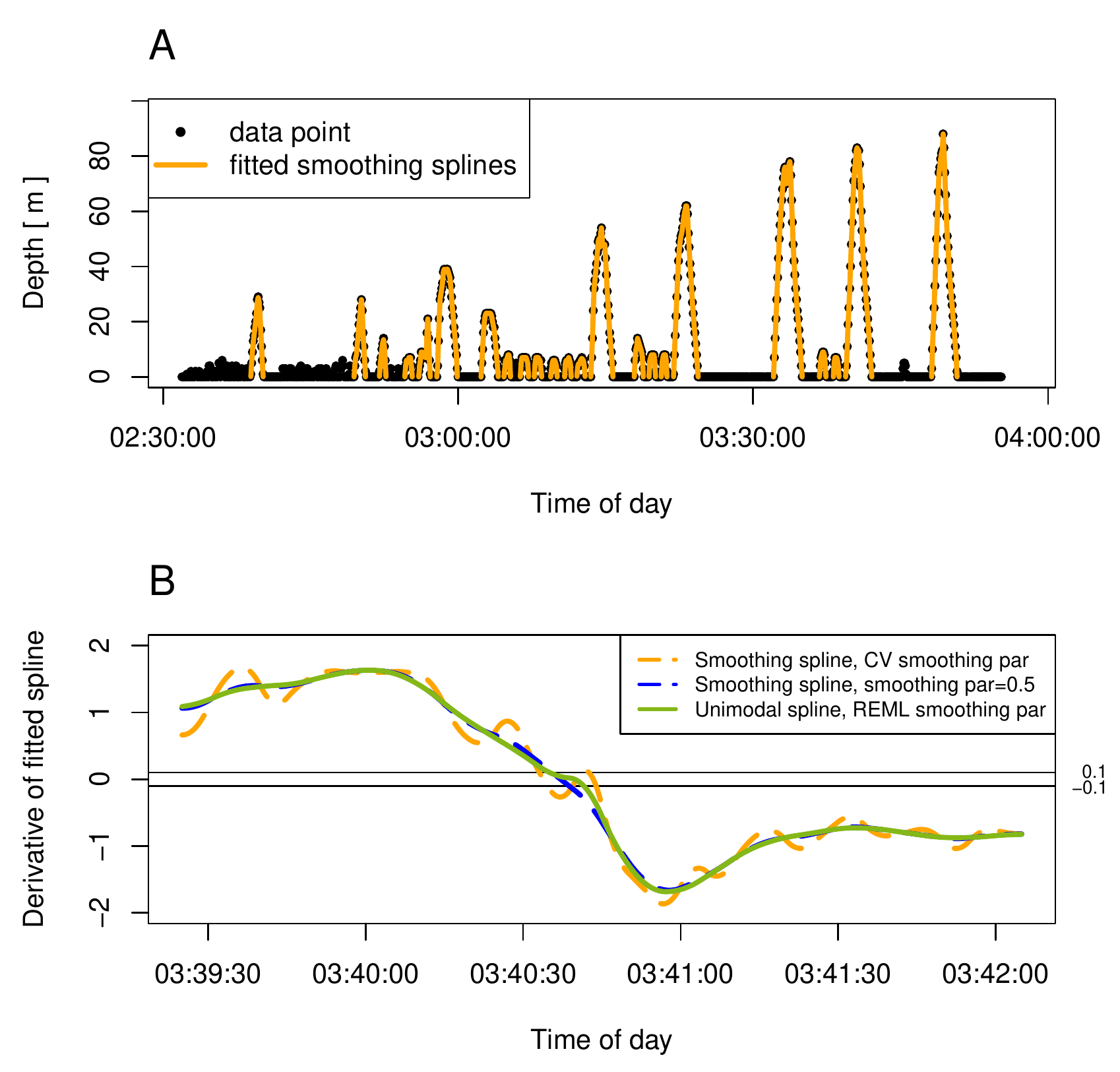}
	\caption{\textbf{Spline regression for the diving depth example.} (A) shows an excerpt from data set \emph{divesTDR} and the smoothing splines fitted with R package \emph{diveMove}. (B) shows derivatives of two fitted smoothing splines (smoothing parameter chosen via cross-validation (CV) and manually) and a unimodal spline (tuning parameter chosen via REML) for the dive second from right in (A). The x-axis is time of day on January 6th 2002 and the y-axis shows the value of the spline derivatives.}\label{fig:dive2}
	\end{minipage}\\
	
If we replace the smoothing spline in the first analysis step by a unimodal spline, in explicit, using piecewise unimodal regression, the derivative has only one sign change and the turning point from descent to ascent is unique, irrespective of the tuning parameter value. We fit a second order difference penalized spline with $q=25$ inner knots and the coefficient vector $\betaf$ and the variance $\sigma^2$ are estimated iteratively, starting with an initial variance estimate of 2 (function \texttt{unireg} with arguments \texttt{abstol = 0.01} and \texttt{sigmasq = 2}). The derivative of the resulting unimodal spline is also shown in Figure \ref{fig:dive2}B.\\
An obvious and welcome side effect of the unimodality constraint is that in contrast to the smoothing spline approach the choice of the tuning parameter has per construction no influence on the uniqueness of the turning point. Tuning can be done via data-driven REML estimation and the user is not confronted with this task.\\
Since the animal needs to come back to the surface to breathe and the time series is divisible into the individual diversely shaped dives, piecewise unimodal regression is a suitable approach for this example. On the one hand, the different shapes of the dives make the application of $L_0$-deco impossible, while there is on the other hand no need to employ a more complex deconvolution model with varying peak shapes since there is no overlap of the unimodal curves.\\
\subsection{Astroparticle physics data analysis}\label{sec:fact}
The loading curves caused by each photon hitting a FACT camera pixel are known to have a unimodal shape. As already noted in the introduction, single and multiple photons can arrive at any time. Employing a deconvolution model is suitable since divisibility into unimodal pieces can be excluded and the measured voltage is an accumulation of several peaks, so called loading curves. It is known that each photon induces the same peak in the voltage, which can roughly be described by the parametric wave of the form in Equation (\ref{eq:wave2}). Thus, we apply parametric $L_0$-deco using the function $g(x|\betaf)=U_0 (1-e^{-\frac{x}{\xi_1}})e^{-\frac{x}{\xi_2}}$. We estimate $\betaf=(U_0, \xi_1, \xi_2)'$ during the iterative algorithm, starting with the parameter estimates derived in \citet{Buss2013}: $\betaf^{(0)}=(17.41,4.745, 31.81)'$. In iteration $k$ the current peak shape $\gf^{(k)}$ is obtained via interim evaluations of $g(x|\hat{\betaf}^{(k-1)})$ at $x=0,\hdots,150$, which is roughly the region with non-zero voltage of the wave.\\
Due to measurement error there exist negative values in the measured voltages. Thus, all values are shifted before model fitting so that the minimal value is zero. The authors of \citet{rooi2011} kindly provided their R implementation of the iterative algorithm for estimation of the $L_0$-deconvolution model. We use our adapted version as described in Section \ref{sec:ident} so that it estimates the parametric and not only a pointwise peak shape. In the original implementation the response values and the peak shape, respectively, are scaled to have a maximum value of one. Since this guarantees identifiability of the model parameters, this preprocessing step is maintained, and by visual inspection we choose a value of $\kappa=0.017$ which is also the default defined in the original code. The result (scaled back to the original voltage range) is shown in Figure \ref{fig:factDecoWave}.\\
\begin{minipage}{\linewidth}
\centering
		\captionsetup{type=figure}
		\includegraphics[width=0.8\textwidth]{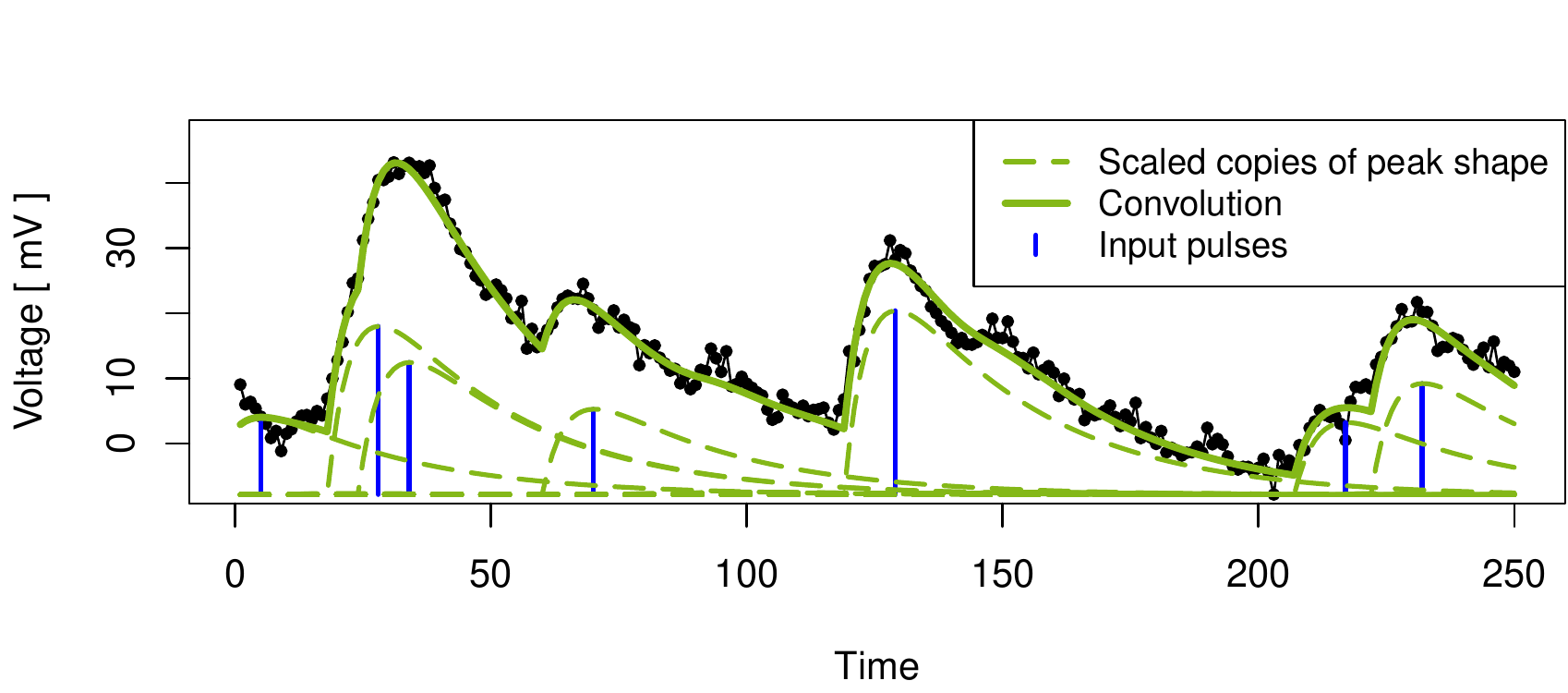}
		\caption{\textbf{FACT time series and fitted deconvolution model.} The individual peaks of the fitted $L_0$-deconvolution model with parametric shapes are given by the dashed green lines and the accumulated signal by the solid green line. The input pulses and their heights are marked with vertical blue bars. The peak shape is modelled with the parametric wave function of form (\ref{eq:wave2}) and each of the seven peaks is a scaled version of it.}\label{fig:factDecoWave}
\end{minipage}\\

Seven input pulses are estimated clearly different from zero and the model represents the data quite well. The parameters of the wave form are estimated slightly different from the starting values as $\hat{U}_0=19.08$, $\hat{\xi}_1=5.66$, $\hat{\xi}_2=28.79$. The resulting wave form is a little higher in the increasing and a little lower in the decreasing part of the peak than the initial peak shape.\\
Again, there is no need to employ the more complex varying $L_0$-deco model. Compared to the diving depth analysis convolution is certainly present in this application, but the individual peaks have the same shape so that the deconvolution model with identical peaks suffices to describe such kind of data. If no information about the wave form had been present, unimodal $L_0$-deco could have been used.
\subsection{Breath gas analysis with IMS}\label{sec:ims}
Each IMS-MCC measurement consists of only a few peaks, which are mostly well-separated (see spectrum A in Figure \ref{fig:data}C). A first approach to model this kind of data is piecewise unimodal regression. We determine three unimodal pieces using a threshold of 50 on the measured voltages and fit separate unimodal splines with ridge penalty and $q=100$ inner knots. We obtain an estimate of the variance from the measurements $y_{36},\hdots,y_{700}$, since it is known that these will definitely contain no peak. The first 35 observations are discarded because the movements of the ion shutter induce some fluctuations in the signal that do not correspond to the usual measurement error of the device. We fix $\sigma^2$ at the estimated value prior to model fitting. The first 700 of the 2499 observations are then discarded for further analyses. In Figure \ref{fig:imsHeur} we see that each of the three peaks is reproduced nicely using this procedure.\\
Problems can occur when the peaks from different molecules are close to each other as, for example, in spectrum B (cf.~Figure \ref{fig:data}D). The second and third peak are so close that their tails overlap. Thus, the intensity measured in between them results from both types of molecules and reflects the accumulation of both concentrations.\\
\begin{minipage}{\linewidth}
\centering
		\captionsetup{type=figure}
		\includegraphics[width=0.8\textwidth]{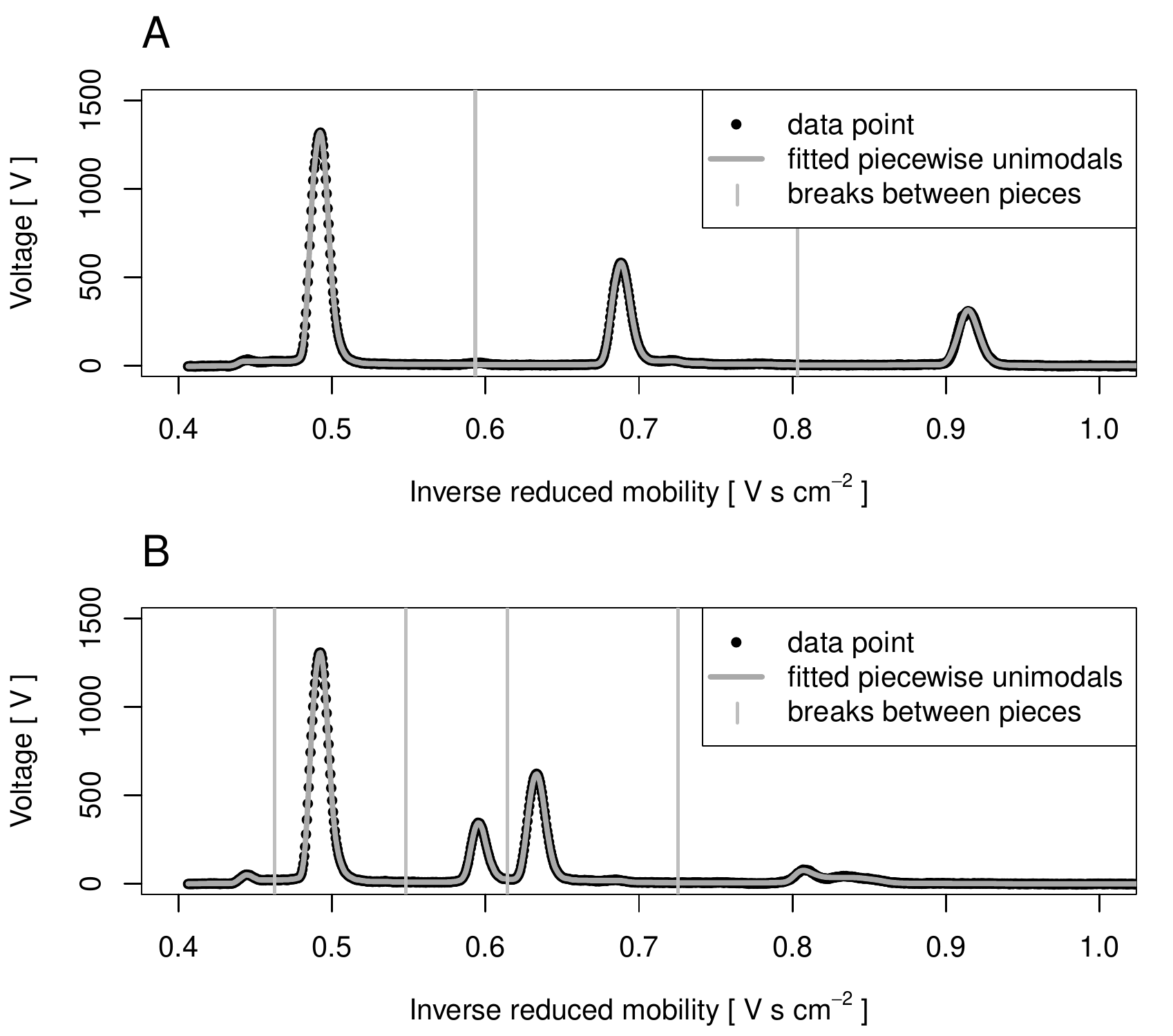}
		\caption{\textbf{Close-ups of IMS spectra A and B with fitted piecewise unimodal regressions.} In both cases, the x-axis was divided into pieces according to a threshold of 50 volt on the intensities. The breaks between the pieces are indicated by vertical lines. A unimodal ridge-penalized spline was fitted to each of the pieces.}\label{fig:imsHeur}
\end{minipage}\\

This cannot be modelled appropriately with piecewise unimodal regression. One might think of employing deconvolution models instead, which are able to handle overlapping as well as non-overlapping peaks. The peaks in IMS data cannot be said to have the same peak shape. \citet{kop2015}, for example, use inverse Gaussian peak shapes and describe each peak with a different set of distribution parameters. Therefore, we propose fitting a deconvolution model with varying peak shapes. This reflects the accumulated intensities of the overlapping peaks as well as the diverse shapes of the peaks. Since the number of peaks is not known a priori, varying $L_0$-deco is preferred to the additive unimodal regression.\\
\begin{minipage}{\linewidth}
\centering
		\captionsetup{type=figure}
		\includegraphics[width=0.8\textwidth]{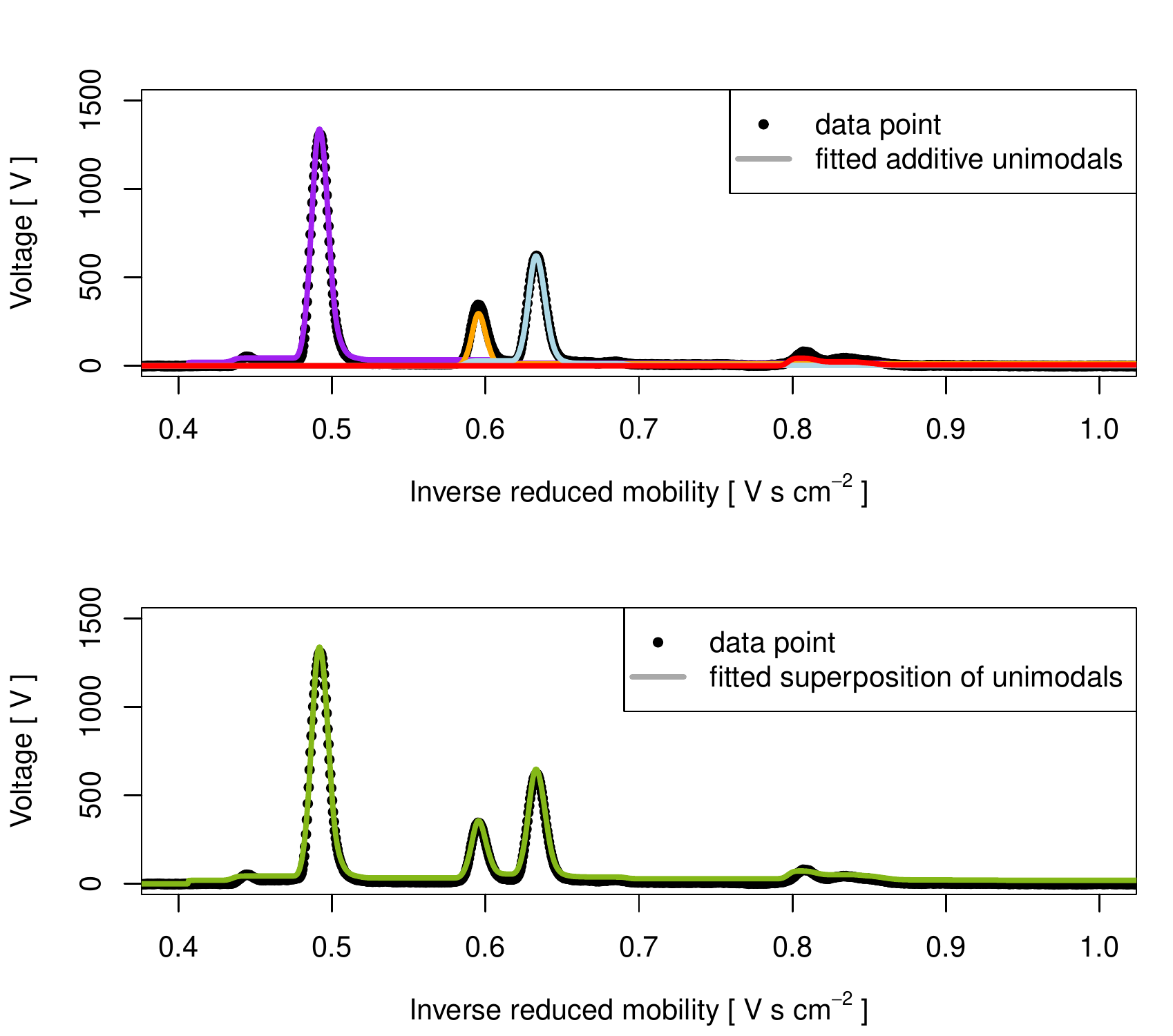}
		\caption{\textbf{Close-up of IMS spectrum B and $L_0$-deconvolution model fit with different peak shapes.} Top: deconvolved peaks of different shapes. Each of the peaks (marked with different colours) is a unimodal spline regression with ridge penalty. Bottom: fitted global function, i.e., the convolution of the above peaks.}\label{fig:imsAddDeco2}
\end{minipage}\\

The variance of the measurement error is estimated as described above. The first 700 observations are discarded afterwards. We use Algorithm ~\ref{alg:deco} (see Appendix) with $q=200$ inner knots, spline degree $k=3$ and $\kappa=0.002$ on the remaining observations. 
Figure \ref{fig:imsAddDeco2} shows the estimated peaks in spectrum B and the fitted global function (scaled back to the original range). As expected, four peaks are modelled with one unimodal spline each. Comparing the convolution model to the piecewise model it is obvious that the two close peaks of spectrum B are much better represented by the deconvolution model.
\section{Summary and Discussion}\label{sec:disc}
In this article we have further developed the unimodal regression introduced in \citet{kbi2014} for use in multimodal applications. We have analysed data from three different application areas -marine biology, astroparticle physics and breath gas analysis- to illustrate that unimodal regression is not only useful when a unimodal relationship between dependent and independent variable is likely, but also as a building block in situations where the relationship is multimodal and has increasing complexity: from piecewise unimodality to accumulations of identically or diversely shaped unimodal functions.\\
Table \ref{tab:choice} summarizes the different data situations and gives recommendations for the model choice. If there is only one mode, the unimodal regression approach by \citet{kbi2014} is directly applicable. Piecewise unimodal regression can be used whenever the underlying process is divisible, in explicit, when there is no overlap between adjacent peaks as, for example, in diving depth data. On the contrary, a deconvolution model is preferable when there is overlap between the unimodal functions and they accumulate to the observed values. Deconvolution with $L_0$-penalty is applicable whenever the peaks have the same shape across the whole data set, as, for example, in the FACT data. If this is not the case, for example, in IMS-MCC data, the deconvolution models with varying peak shapes should be used. Due to an increased computational burden for the additive model we recommend it only in applications where the number of peaks is known. In all other cases the varying $L_0$-deco approach is the method of choice and provides estimates of the number of peaks, their heights and shapes.\\
In comparison to parametric models as, for example, the shifted inverse Gaussian distribution for IMS data, spline regression is a very flexible tool. There is no need to restrict the possible functional relationships to some parametric family, while prior knowledge about the shape can still be incorporated by using a shape constraint (here: unimodality) and/or by a penalty (ridge penalty or penalizing against a parametric function). Another nice characteristic of splines is the simplicity of calculating derivatives and that the derivatives of shape constrained splines also "inherit" shape properties. In the case of the marine biology data the monotonicity of the first derivative simplifies the subsequent analyses, namely the detection of descent and ascent phase of the dive by finding the zero of the derivative.\\
The analyses in this article provide an indication for the usefulness of unimodal regression in the presented and in further applications. Of course there are situations where unimodality is not as likely as, for example, in the case of the peaks in IMS data or the FACT loading curves. Actually, one could argue that the dive of a marine animal is not strictly unimodal since the animal might also descend to a certain depth, make some smaller upward and downward movements and then ascend to the surface again. Those wiggles at the bottom are flattened out by the unimodal spline approach and the turning point that fits the data best divides the dive into descent and ascent. Thus, the unimodal model may be regarded as a simplification of a dive, but it is a suitable one and provides an automated estimation process for biological scientists. Adaptations of the method to aims other than the division into the two phases are also possible.\\
So far, all $L_0$-deconvolution models have been tuned manually and there is still need to explore methods for automatic choices of the tuning parameter. For example, one can also think of a Bayesian model formulation, where sparsity of the input pulses can be achieved by spike-and-slab priors.\\
Additionally, subsequent analysis steps like classification or integration are often common in the described situations. Hence, systematic evaluations of the impact of the modelling step on the final outcome are needed. Such performance studies could be conducted, for example, along the lines of \citet{Hausch2013}.
\section*{Acknowledgements}
Part of the work on this article has been supported by Deutsche Forschungsgemeinschaft (DFG) within the Collaborative Research Center SFB 876 "Providing Information by Resource-Constrained Analysis", project C4. We would like to thank SFB 876 projects TB1 and C3 and the FACT collaboration for providing the data examples (TB1: IMS-MCC data, C3 and FACT collaboration: FACT data) and for their appreciated comments on the here presented research. We are grateful to Sebastián P. Luque for the helpful discussion about the application of our methods to the diving depth data. Many thanks also to Paul H.C. Eilers for very valuable comments, which led to important improvements of the manuscript, and to Johan J. de Rooi for providing the R code for estimation of the deconvolution model with $L_0$-penalty.
\bibliography{mybib6authors}

\begin{thebibliography}{}

\bibitem[Anderhub et~al., 2013]{fact2013}
Anderhub, H., Backes, M., Biland, A., Boccone, V., Braun, I., Bretz, T.,
  Bu\ss\, J., Cadoux, F., Commichau, V., Djambazov, L., Dorner, D., Einecke,
  S., Eisenacher, D., Gendotti, A., Grimm, O., von Gunten, H., Haller, C.,
  Hildebrand, D., Horisberger, U., Huber, B., Kim, K.-S., Knoetig, M.~L.,
  Koehne, J.-H., Kraehenbuehl, T., Krumm, B., Lee, M., E.Lorenz, Lustermann,
  W., Lyard, E., Mannheim, K., Meharga, M., Meier, K., Montaruli, T., Neise,
  D., Nessi-Tedaldi, F., Overkemping, A.-K., Paravac, A., Pauss, F., Renker,
  D., Rhode, W., Ribordy, M., Roeser, U., Stucki, J.-P., Schneider, J.,
  Steinbring, T., Temme, F., Thaele, J., Tobler, S., Viertel, G., Vogler, P.,
  Walter, R., Warda, K., Weitzel, Q., and Zaenglein, M. (2013).
\newblock Design and operation of {FACT} - the first {G-APD} {C}herenkov
  telescope.
\newblock {\em Journal of Instrumentation}, 8(06):P06008.

\bibitem[Biland et~al., 2014]{Bil2012}
Biland, A., Bretz, T., Bu\ss\, J., Commichau, V., Djambazov, L., Dorner, D.,
  Einecke, S., Eisenacher, D., Freiwald, J., Grimm, O., von Gunten, H., Haller,
  C., Hempfling, C., Hildebrand, D., Hughes, G., Horisberger, U., Knoetig,
  M.~L., Kr\"ahenb\"uhl, T., Lustermann, W., Lyard, E., Mannheim, K., Meier,
  K., Mueller, S., Neise, D., Overkemping, A.~K., Paravac, A., Pauss, F.,
  Rhode, W., R\"oser, U., Stucki, J.~P., Steinbring, T., Temme, F., Thaele, J.,
  Vogler, P., Walter, R., and Weitzel, Q. (2014).
\newblock Calibration and performance of the photon sensor response of {FACT} -
  the first {G-APD} {C}herenkov telescope.
\newblock {\em Journal of Instrumentation}, 9(10):P10012.

\bibitem[B\"odeker and Baumbach, 2009]{Boed2009}
B\"odeker, B. and Baumbach, J.~I. (2009).
\newblock Analytical description of ims-signals.
\newblock {\em International Journal for Ion Mobility Spectrometry},
  12(3):103--108.

\bibitem[Bu\ss, 2013]{Buss2013}
Bu\ss, J.~B. (2013).
\newblock Fact - signal calibration: Gain calibration and development of a
  single photon pulse template for the fact camera.
\newblock Diploma thesis, Faculty of Physics, TU Dortmund University.

\bibitem[{de Rooi} and Eilers, 2011]{rooi2011}
{de Rooi}, J. and Eilers, P. (2011).
\newblock Deconvolution of pulse trains with the {$L_0$} penalty.
\newblock {\em Analytica Chimica Acta}, 705:218--226.

\bibitem[{de Rooi} et~al., 2014]{rooi2014}
{de Rooi}, J., Ruckebusch, C., and Eilers, P. (2014).
\newblock Sparse deconvolution in one and two dimensions: Applications in
  endocrinology and single-molecule fluorescence imaging.
\newblock {\em Analytical Chemistry}, 86:6291--6298.

\bibitem[Dierckx, 1993]{Die1993}
Dierckx, P. (1993).
\newblock {\em Curve and surface fitting with splines}.
\newblock Oxford Science Publications. Clarendon Press, Oxford.

\bibitem[Ei\-lers, 2005]{Eilers2005}
Ei\-lers, P. H.~C. (2005).
\newblock Unimodal smoothing.
\newblock {\em Journal of Chemometrics}, 19:317--328.

\bibitem[Ei\-lers and Marx, 1996]{EilMar1996}
Ei\-lers, P. H.~C. and Marx, B.~D. (1996).
\newblock Flexible smoothing with {B}-splines and penalties.
\newblock {\em Statistical Science}, 11(2):89--121.

\bibitem[Fris\'{e}n, 1986]{Fri1986}
Fris\'{e}n, M. (1986).
\newblock Unimodal regression.
\newblock {\em The Statistician}, 35:479--485.

\bibitem[Gr\"un and Leisch, 2008]{Gruen2008}
Gr\"un, B. and Leisch, F. (2008).
\newblock Flexmix version 2: Finite mixtures with concomitant variables and
  varying and constant parameters.
\newblock {\em Journal of Statistical Software}, 28(4):1--35.

\bibitem[Halsey et~al., 2007]{Hals2007}
Halsey, L.~G., Bost, C.-A., and Handrich, Y. (2007).
\newblock A thorough and quantified method for classifying seabird diving
  behaviour.
\newblock {\em Polar Biology}, 30:991--1004.

\bibitem[Hastie et~al., 2009]{Hast2009}
Hastie, T., Tibshirani, R., and Friedman, J. (2009).
\newblock {\em The Elements of Statistical Learning: Data Mining, Inference,
  and Prediction.}
\newblock Springer Series in Statistics. Springer, second edition.

\bibitem[Hauschild et~al., 2013]{Hausch2013}
Hauschild, A.-C., Kopczynski, D., D'Addario, M., Baumbach, J.~I., Rahmann, S.,
  and Baumbach, J. (2013).
\newblock Peak detection method evaluation for ion mobility spectrometry by
  using machine learning approaches.
\newblock {\em Metabolites}, 3:277--293.

\bibitem[K\"ollmann, 2016]{unireg2016}
K\"ollmann, C. (2016).
\newblock {\em uniReg: Unimodal penalized spline regression using B-splines}.
\newblock R package version 1.1.

\bibitem[K\"ollmann et~al., 2014]{kbi2014}
K\"ollmann, C., Bornkamp, B., and Ickstadt, K. (2014).
\newblock Unimodal regression using {B}ernstein-{S}choenberg-splines and
  penalties.
\newblock {\em Biometrics}, 70:783--793.
\newblock {d}oi: 10.1111/biom.12193.

\bibitem[Kopczynski et~al., 2012]{kop2012}
Kopczynski, D., Baumbach, J.~I., and Rahmann, S. (2012).
\newblock Peak modeling for {I}on {M}obility {S}pectrometry measurements.
\newblock In {\em Proceedings of the 20th European Signal Processing Conference
  (EUSIPCO)}, pages 1801--1805.

\bibitem[Kopczynski and Rahmann, 2015]{kop2015}
Kopczynski, D. and Rahmann, S. (2015).
\newblock An online peak extraction algorithm for ion mobility spectrometry
  data.
\newblock {\em Algorithms for Molecular Biology}, 10(1):1--14.

\bibitem[Luque, 2007]{diveMove2007}
Luque, S.~P. (2007).
\newblock Diving behaviour analysis in {R}.
\newblock {\em {R News}}, 7(3):8--14.
\newblock Contributions from: J. P. Y. Arnould, L. Dubroca, and A. Liaw.

\bibitem[Luque and Fried, 2011]{Luq2011}
Luque, S.~P. and Fried, R. (2011).
\newblock Recursive filtering for zero offset correction of diving depth time
  series with {GNU} {R} package dive{M}ove.
\newblock {\em PLoS ONE}, 6(1):e15850.

\bibitem[Oller-Moreno et~al., 2015]{Oller2015}
Oller-Moreno, S., Singla-Buxarrais, G., Jiménez-Soto, J., Pardo, A.,
  Garrido-Delgado, R., Arce, L., and Marco, S. (2015).
\newblock Sliding window multi-curve resolution: Application to gas
  chromatography-ion mobility spectrometry.
\newblock {\em Sensors and Actuators B: Chemical}, 217:13 -- 21.
\newblock Selected Papers from the 15th International Meeting on Chemical
  Sensors, 16-19 March 2014, Buenos Aires, Argentina.

\bibitem[Pomareda et~al., 2010]{Pomareda2010}
Pomareda, V., Calvo, D., Pardo, A., and Marco, S. (2010).
\newblock Hard modeling multivariate curve resolution using lasso: Application
  to ion mobility spectra.
\newblock {\em Chemometrics and Intelligent Laboratory Systems}, 104(2):318 --
  332.

\bibitem[{R Core Team}, 2016]{R2016}
{R Core Team} (2016).
\newblock {\em R: A Language and Environment for Statistical Computing}.
\newblock R Foundation for Statistical Computing, Vienna, Austria.

\bibitem[Rossoni and Feng, 2006]{Ross2006}
Rossoni, E. and Feng, J. (2006).
\newblock A nonparametric approach to extract information from interspike
  interval data.
\newblock {\em Journal of Neuroscience Methods}, 150:30--40.

\bibitem[Tauler, 1995]{Tauler1995}
Tauler, R. (1995).
\newblock Multivariate curve resolution applied to second order data.
\newblock {\em Chemometrics and Intelligent Laboratory Systems}, 30(1):133 --
  146.
\newblock InCINC '94 Selected papers from the First International Chemometrics
  Internet Conference.

\bibitem[Vogtland and Baumbach, 2009]{Vogt2009}
Vogtland, D. and Baumbach, J.~I. (2009).
\newblock Breit-{W}igner-{F}unction and {IMS}-signals.
\newblock {\em International Journal for Ion Mobility Spectrometry},
  12:109--114.

\bibitem[Westhoff et~al., 2009]{West2009}
Westhoff, M., Litterst, P., Freitag, L., Urfer, W., Bader, S., and Baumbach,
  J.~I. (2009).
\newblock Ion mobility spectrometry for the detection of volatile organic
  compounds in exhaled breath of lung cancer patients.
\newblock {\em Thorax}, 64:744--748.
\newblock April 2012.

\end{thebibliography}
\bibliographystyle{apalike}
\section*{Appendix}
\begin{algorithm}[ht]
 \KwIn{$\xf$, $\yf$, $q$, $k$, $\kappa$}
 \KwOut{$\hat{\yf}$, $\hat{\af}$, $\hat{\Gf}$}
 \Begin(Initialization){
 $\yf := u(\yf)$; $d := q+k+1$; $\beta := 10^{-5}$\;
 \For{$j=1,\hdots,d$}{
 		fit spline $g_j$ with fixed mode $j$ to $\yf$\; $\Gf_{.j} := u(g_j(x_1),\hdots,g_j(x_n))$\;
 		}
 $\mathbf{W} := \kappa \cdot \mathcal{I}_d$; $\af := \left(\Gf' \Gf + \mathbf{W} \right)^{-1} \Gf' \yf$\;
 $\mathbf{W} := \mathit{diag}\left(\frac{1}{a_j^2 + \beta^2}\right)_{j=1,\hdots,d}$; $\af := \left(\Gf' \Gf + \mathbf{W} \right)^{-1} \Gf' \yf$\;
\If{$a_j<0.0001$}{$a_j :=0 $}
}
\Repeat{$\max_j |a_j^{(old)} - a_j|<0.001$}{
$\af^{(old)} := \af$\; $\mathcal{L} := \bigcup_{\{\ell: a_{\ell}\ne 0\}} \{\ell-2,\hdots,\ell+2\} \cap \{1,\hdots,d\}$\;
	\For{$j \in \mathcal{L}$}{
	$\tilde{\af} := (a_1,\hdots,a_{j-1},0,a_{j+1},\hdots,a_d)$\; $\tilde{\yf} := \yf - \Gf \tilde{\af}$\;
	fit spline $g_j$ with fixed mode $j$ to $\tilde{\yf}$\; $\Gf_{.j} := u(g_j(x_1),\hdots,g_j(x_n))$\;
	\For{$i=1,\hdots,5$}{
 		$\mathbf{W} := \mathit{diag}\left(\frac{1}{a_j^2 + \beta^2}\right)_{j=1,\hdots,d}$\; $\af := \left(\Gf' \Gf+ \kappa \mathbf{W} \right)^{-1} \Gf' \yf$\;
 		}
		\If{$a_j<0.0001$}{$a_j :=0 $}
		}
}
\vspace{0.2cm}
    \caption[]{Algorithm for deconvolution with $L_0$-penalty and varying peak shapes.\\ The diagonal matrix is denoted by $\mathit{diag}$ and the function $\protect u: \R^n \to [0,1]^n$ is given by \\ $\protect u(\zf)=
		\begin{cases}
		\frac{\zf - \min(\zf)}{\max(\zf) - \min(\zf)}, & \min(\zf)<\max(\zf)\\
		\frac{\zf}{\max(\zf)}, & \min(\zf)=\max(\zf)
		\end{cases}$.}\label{alg:deco}
\end{algorithm}
\end{document}